**Magnetic properties, chemical and phase composition of thin manganese silicide films fabricated by pulsed laser deposition**

M.V. Dorokhin, Yu.M. Kuznetsov, V.P. Lesnikov, A.V. Kudrin, I.V. Erofeeva, A.V. Boryakov, R.N. Kryukov, D.E. Nikolitchev, S.Yu. Zubkov, V.N.Trushin, P.B. Demina

Lobachevsky State University of Nizhny Novgorod, 603950, Gagarin ave.23, Nizhniy Novgorod, Russia

dorokhin@nifti.unn.ru

Manganese silicide ($Mn_xSi_y$) thin films with Mn content ($C_{Mn}$) varied from 24 at. % to 52 at. % were grown on i-GaAs (100) substrates. Chemical, phase composition and room temperature magnetic properties of the films were investigated. It was demonstrated that manganese silicide films revealing ferromagnetic properties are multiphase systems. The phases, which are believed to be non-magnetic at room temperature, are compounds with Mn concentration exceeding the average Mn content of the film (higher manganese silicide, MnSi and $Mn_5Si_3$). Magnetic properties revealed are attributed with the secondary phases which are Mn doped silicon (Mn:Si) or Mn depleted compounds. For nearly single phase films (with very small 2-nd phase ratio) no room temperature ferromagnetism was observed.

# 1. Introduction

Manganese silicide-based thin films and nanostructures have been intensively studied within the last two decades [1-18]. Great attention paid to the investigation of these films is due to a number of practical applications such as skyrmion magnetic memory elements [1-3], thermoelectric energy converters [4,5] and ferromagnetic elements of the silicon spintronics [6-18]. For the latter $Mn_xSi_y$ films with 0.8-50 at. % of Mn are used as a source of spin-polarized carriers for spin manipulation in silicon-based spintronic devices [19]. From this point of view, it is of extreme importance to fabricate the material possessing ferromagnetic properties above room temperature: that would provide spintronic device operation at least at 300 K. At first glance, most of known manganese silicide phases are either low temperature ferromagnetic with Curie point typically below 300 K or antiferromagnetic [19-26]. This statement applies to doped Si:Mn with low Mn content [15,20], higher manganese silicide (HMS) [21-23], manganese monosilicide (MnSi) [24,25] and even $Mn_5Si_3$ [23,26]. Surprisingly much wider number of room temperature ferromagnetic manganese silicide structures was obtained experimentally. There are numerous reports on the room temperature ferromagnetism in such systems including different manganese silicide phases and having different Mn content [12,13,16,18,27-30]. The experimental results were followed by theoretical considerations that suggested a set of mechanisms to describe ferromagnetic ordering in thin $Mn_xSi_y$ films. In [31] the calculations of electron structure of diluted Si:Mn have been performed and hole-mediated RKKY mechanism for ferromagnetic ordering was proposed. Some papers show the formation of higher manganese silicide precipitates in Si:Mn films and relatively low magnetic moment was attributed with these precipitates [13,32-34]. In [11] ferromagnetism of diluted Mn-doped silicon (Si:Mn) structures was discussed. It was suggested that the complex of two Mn-Mn interstitials and one Mn substitutional atom compose a ferromagnetically ordered state. The attempt to generalize some of experimental results was made in [35] where a phenomenological model describing ferromagnetic ordering of $Mn_xSi_y$ system was developed.

Abovementioned ambiguities in $Mn_xSi_y$ thin films characterization are evidently related with strong sensitivity of the properties of these films to parameters of a growth process. Moreover, grown films very often turn to be multiphase systems with phase composition being strongly dependent on both the method for manganese silicide fabrication and Mn content [13-15, 30,32,33].

In the present paper, we report on structural, chemical and magnetic investigation of $Mn_xSi_y$ thin films grown on i-GaAs (100) substrates by pulsed laser deposition technique. The Mn content in the films was varied between 24 at. % and 52 at. %. The peculiarities of room temperature ferromagnetism of these layers are discussed through the analysis of their phase composition. It was suggested that magnetic properties at 300 K are provided by a phase of Mn-doped silicon. Samples with a low content of Si:Mn phase are characterized by rather low magnetization value or even show no magnetic properties at 300 K (for the lowest Si:Mn phase content).

## 2. Experimental techniques

The $Mn_xSi_y$ layers with different Mn content were grown on semi-insulating GaAs (100) substrates by pulsed laser deposition of semiconducting Si and metallic Mn targets in a vacuum chamber with a background gas pressure of about $2\times10^{-5}$ Pa. The growth temperature ($T_g$) in these experiments was 350 °C. The Mn content was first set by technological parameter $Y_{Mn}=t_{Mn}/(t_{Mn}+t_{Si})$, where $t_{Mn}$ and $t_{Si}$ are the ablation times of Mn and Si targets, respectively. More precise value of Mn content in the deposited $Mn_xSi_y$ film was afterwards determined by energy dispersive X-ray spectroscopy (EDX) using a JSM-IT300LV (JEOL) scanning electron microscope with X-Max[N] 20 (Oxford Instruments) detector. Analysis of the layer composition was carried out using silicon $K\alpha$ and manganese $K\alpha$ characteristic X-ray radiation lines; signal from GaAs substrate was excluded. Accelerating voltage was 20 kV; current of the electron probe did not exceed 0.5 nA. The thickness of all $Mn_xSi_y$ samples was technologically set equal to 40 nm. Control of the thickness was carried out by selective etching of the part of $Mn_xSi_y$ film and by further measuring step height on the border of etched and non-etched parts using atomic force microscope (Fig.1a). Determined thickness variation from sample to sample did not exceed 5 nm. Full list of measured parameters of the samples is given in Table 1.

The crystalline structure of deposited films was studied by X-ray diffraction (XRD) technique, using a D8 Discover diffractometer (Bruker). $CuK\alpha_1$ X-ray beam of with a focal-spot size of $0.1 \times 10$ mm$^2$ was limited by a slit with a diameter of 1 mm. The size of illumination area on the surface of single-crystal wafer in this case was $1.2\times1.8$ mm$^2$.

Table 1. The list of investigated samples with measured parameters

| # | $C_{Mn}$ (EDX), at. % | $M_{rel}$, $10^{-6}$ emu/cm$^2$ | $M_{av}$, $\mu_b$/Mn | $C_{Mn}$ - (XPS), at. % | $C_{Si-Mn}$ - at. % ($\eta$) |
|---|---|---|---|---|---|
| 1 | 23 | 131 | 0.34 | 26.8 | 58.2 (2.18) |
| 2 | 29 | 41.5 | 0.14 | 29.3 | 54.5 (1.86) |
| 3 | 33 | 0 | 0 | 34.7 | 56.6 (1.63) |
| 4 | 44 | 183 | 0.26 | 47.4 | 52.6 (1.11) |
| 5 | 52 | 187 | 0.22 | 49.8 | 50.2 (1.01) |

where $C_{Mn}$ – Mn fraction in $Mn_xSi_y$ film, $M_{rel}$ – relative magnetization, $M_{av}$ – average magnetic moment, $C_{Si-Mn}$ – fraction of Si atoms bound to Mn atoms, $\eta=C_{Si-Mn}/C_{Mn}$.

Along with EDX, the distribution over the depth of constituent elements concentration in $Mn_xSi_y$ films was also measured by means of X-ray photoelectron spectroscopy (XPS) based on Multiprobe RM instrument (Omicron Nanotechnology GmbH). Al $K_\alpha$ radiation with an energy of 1486.7 eV was used to excite photoemission. Diameter of analysis area was 3 mm. Layer profiling over the sample depth was carried out by sputtering with Ar$^+$ ions. Ion beam had 20 mm diameter, 1 keV energy, uniform radial distribution of ion current density and 45° angle of inclination. The analysis of peak shift within this method allows one to draw the conclusions on chemical bonds and phase composition of the films.

Atomic concentration in the layers was determined by the method of relative sensitivity factors [36]. Silicon $2p$, Mn $2p_{3/2}$, O $1s$ and C $1s$ photoelectron lines (PE lines) were analyzed. Structure of PE lines due to chemical shifts was investigated using mathematical software CasaXPS [37]. The presence of a chemical bond between manganese and silicon was determined during spectral analysis of Si $2p$ PE line (Fig. 1b). After eliminating effect of surface charging using calibration with C $1s$ and O $1s$ PE lines, Si $2p$ peak was approximated by the curve (Fig.1b, curve 1) with parameters corresponding to those, which were obtained by analyzing silicon from Si substrate. The function originally had a normal distribution, and variable parameters were the energy position, FWHM, and degree of asymmetry. Then, a second component was added (Fig.1b, curve 2), with parameters borrowed from [38,39]. The intensities of these two curves were programmed to values at which maximum coincidence of their envelope with experimental PE line was achieved. By the ratio of the areas of these two curves, the ratio of silicon concentrations in Si-Si and Si-Mn bonds was found.

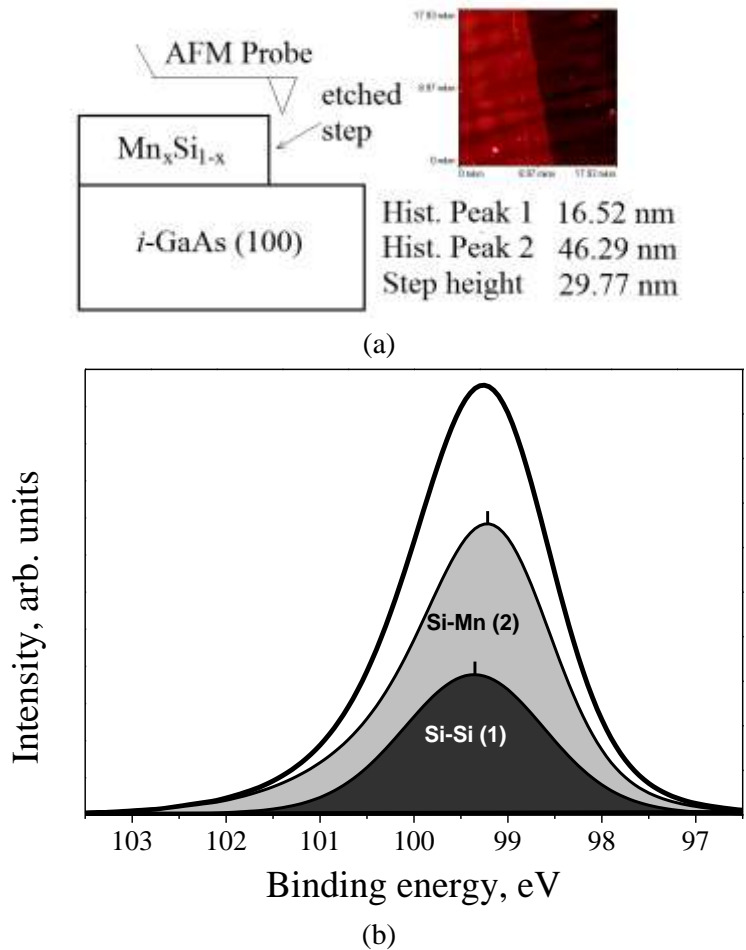

Fig. 1. (a) The scheme of the sample etched-step measurement of thickness; (b) Si 2*p* peak approximated by PE lines obtained from silicon substrate and from literature data.

The magnetic properties of $Mn_xSi_y$ films were studied at room temperature via magnetization vs magnetic field measurements, which were carried out by means of a calibrated alternating-field gradient magnetometer [40]. The magnetic field with maximum value of 1700 Oe was applied parallel to the surface of the film (in-plane geometry). For magnetization measurements, part of the sample with lateral size of approximately 2-3×5 mm was used.

## 3. Experimental results

Magnetic field dependencies of magnetization of the films measured at 300 K for all investigated samples are shown at Fig.2. Diamagnetic contribution of i-GaAs substrate was subtracted from magnetization values of all samples. The latter was approximated as linear function of the magnetic field and calculated separately for each *M*(*H*) curve. The example of GaAs substrate diamagnetic signal is shown at Fig.2f. Taking substrate contribution into account one can see that all investigated samples demonstrate nonlinear *M*(*H*) dependencies (with only one exception that will be discussed below). Hysteresis loop at *M*(*H*) dependencies was revealed for samples 1, 4 and 5. It is important to note that hysteretic *M*(*H*) behavior is the evidence of measured samples possessing ferromagnetic properties at room temperature [16,27]. For sample 2 no hysteresis loop was detected which is supposedly due to relatively low magnetization signal (the experimental error value is greater than the remnant magnetization). In sample 3 the magnetic signal is also relatively low and does not exceed the instrumental error. The latter was obtained by performing magnetization measurements in magnetometer chamber without a sample. Supposedly, it can be attributed with some uncontrolled magnetic particles, which can remain in magnetometer since previous measurements. Taking into account such an instrumental error, we will consider magnetization value of sample 3 being equal to zero.

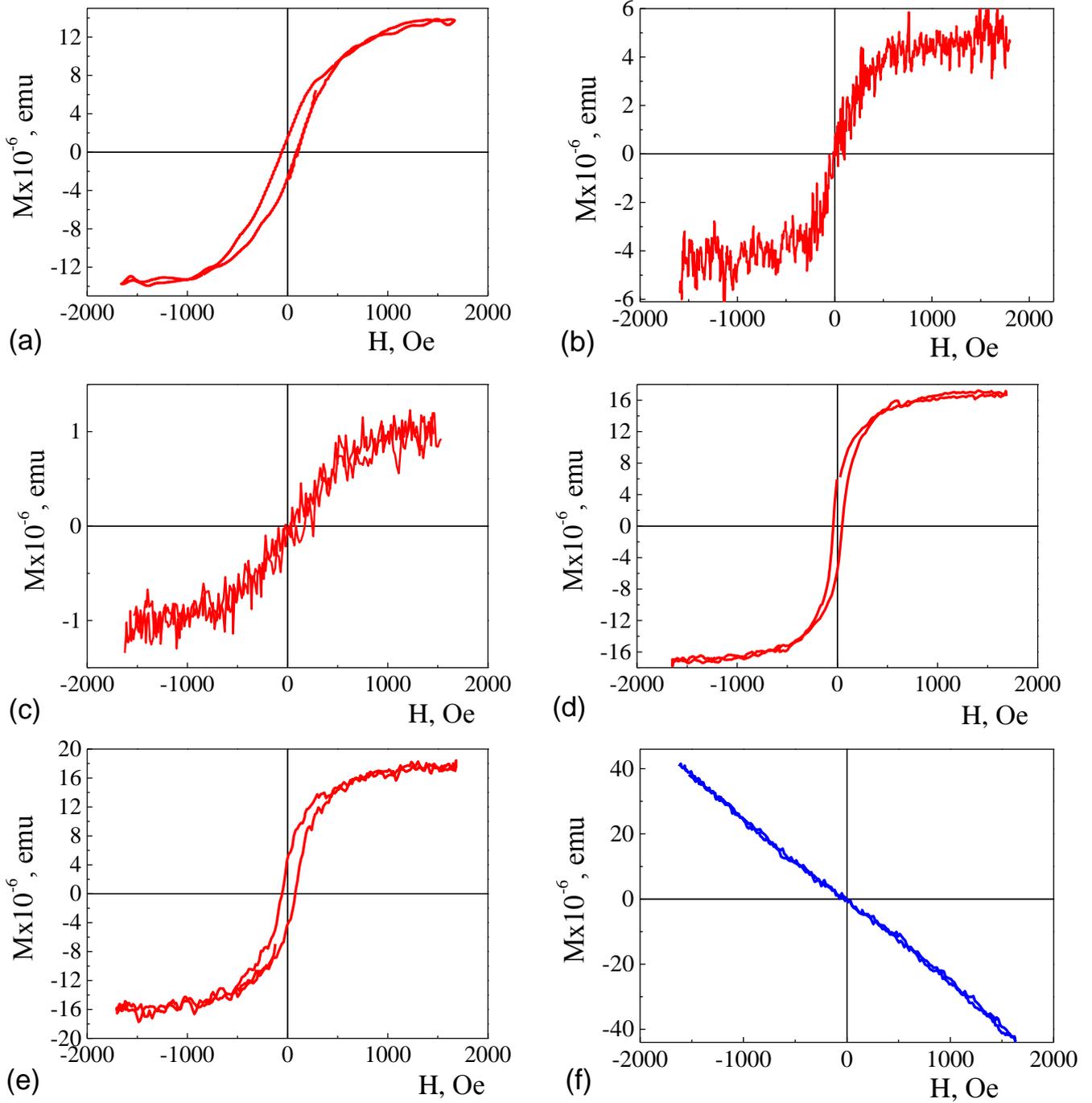

Fig.2 Magnetic field dependencies of magnetization of investigated samples measured at 300 K: (a) – sample 1, (b) – sample 2, (c) – sample 3, (d) – sample 4, (e) – sample 5, (f) diamagnetic signal from thick GaAs substrate.

The following analysis of magnetic properties is focused on differences in saturation magnetization values for each of the investigated samples. Although calibration of magnetometer may introduce some ambiguity into definition of the absolute value of magnetization, saturation magnetization per unit area for different samples can be compared with one another (provided thickness of the films was always the same). Corresponding data with considering instrumental error value is given in Table 1. Nonmonotonic dependence on Mn content was found: magnetization value for sample with the lowest Mn content of 24 at. % was as high as $1.31 \times 10^{-4}$ emu/cm$^2$. Increasing Mn concentration only by 5 atomic % to 29 at. % leads to a 3-fold drop of the saturation magnetization. With further increase of Mn concentration (to 33 at. %) saturation magnetization value drops down to zero. In samples with Mn concentration close to 50 at. % magnetic properties were revealed at 300 K and even the highest magnetization values were obtained.

Using data in Table 1 and measuring geometrical parameters of the sample one can find the average magnetic moment per Mn atom ($M_{ave}$ - in $\mu_b$/Mn, where $\mu_b$ is Bohr magneton). We emphasize that the data obtained is only a rough estimation of the average magnetic moment. Exact value of Mn atoms in samples is different for various manganese silicide phases (substitutional doping [11,14], interstitial doping [11], higher manganese silicide [22,41] or manganese monosilicide [24,42]). However, the estimated difference in number of Mn atoms for each compound is not higher than 30 % which cannot account for the revealed difference in the average magnetic moment.

Let us first focus on discussion of the magnetic properties of low-Mn content films. Despite relatively small difference between Mn content in samples 1-3 the average magnetic moment per Mn atom varies significantly from sample to sample. This draws us to conclusion that the major factor, which influences magnetization values in $Mn_xSi_y$ films, is not the Mn concentration but the abovementioned phase composition of the film (Si doped with Mn, higher manganese silicide, manganese monosilicide, etc.). Within this assumption, it is the different phase composition in $C_{Mn}$ = 24 %, 29 % and 33 % films that governs magnetic ordering of Mn atoms. This conclusion can be extended to samples with nearly 50 % Mn content, with only proviso that the film composition differs.

Further support of our conclusion is the fact that most of the room temperature ferromagnetic manganese silicide phases are characterized by much greater value of relative magnetic moment as compared with the investigated samples. This is in a good agreement with the assumption that investigated samples include at least two manganese silicide phases one of which is non-magnetic or weak magnetic phase at 300 K that decreases the average $\mu_b$/Mn value.

## 4. Structural investigations and element analysis

In order to support the conclusion drawn in previous section the investigations of samples crystalline structure and chemical phase composition were carried out by means of X-Ray diffraction measurements. The XRD spectra are plotted at Fig.3. All of the investigated samples demonstrate rather poor crystalline quality. Dominating diffraction peak is a background peak of [200] GaAs substrate at 2Θ≈31.7° (not shown). Subtracting background one can see a set of small peaks corresponding to different phases of manganese silicide. In samples 1-3 the higher manganese silicide phase was detected ($Mn_{27}Si_{47}$ for sample 1 and $Mn_{15}Si_{26}$ for samples 2 and 3).

In sample 4 with ~44 at. % of Mn the manganese monosilicide phase was revealed on X-Ray diffraction spectrum (Fig.3b, Curve 4). The Mn-related peak, which is present at the diffraction spectrum of sample 4, is attributed with a Mn droplet that could be deposited on substrate surface from laser plasma. Droplet formation on the substrate is typical for pulsed laser deposition technique. In further analysis, the presence of this droplet can be disregarded since it should not influence ferromagnetic properties. For the sample 5 with 52 at. % of Mn the peaks related to $Mn_5Si_3$ phase were revealed. Small amplitude of the peaks evidences again that film is highly disordered.

We note that in all cases the concentration of Mn in dominating manganese silicide compound is higher than the average Mn concentration of the film: in samples 1-3 Mn content of ~24 to 33 at. % is below bulk higher manganese silicide value (~ 36 at. %); in sample 4 Mn content is 44 at. % which is below MnSi value (50 %); finally, Mn content in sample 5 (52 at. %) is below the $Mn_5Si_3$ value (62,5 at. %). Thus one can suggest the co-existence of the phases with lower Mn concentration (in particular, Si:Mn – Mn-doped silicon phase). The absence of peaks related with silicon and other phases at XRD spectra can be attributed with smaller fraction of 2-nd phase as compared with the dominating one and with not well enough crystalline quality of the entire film.

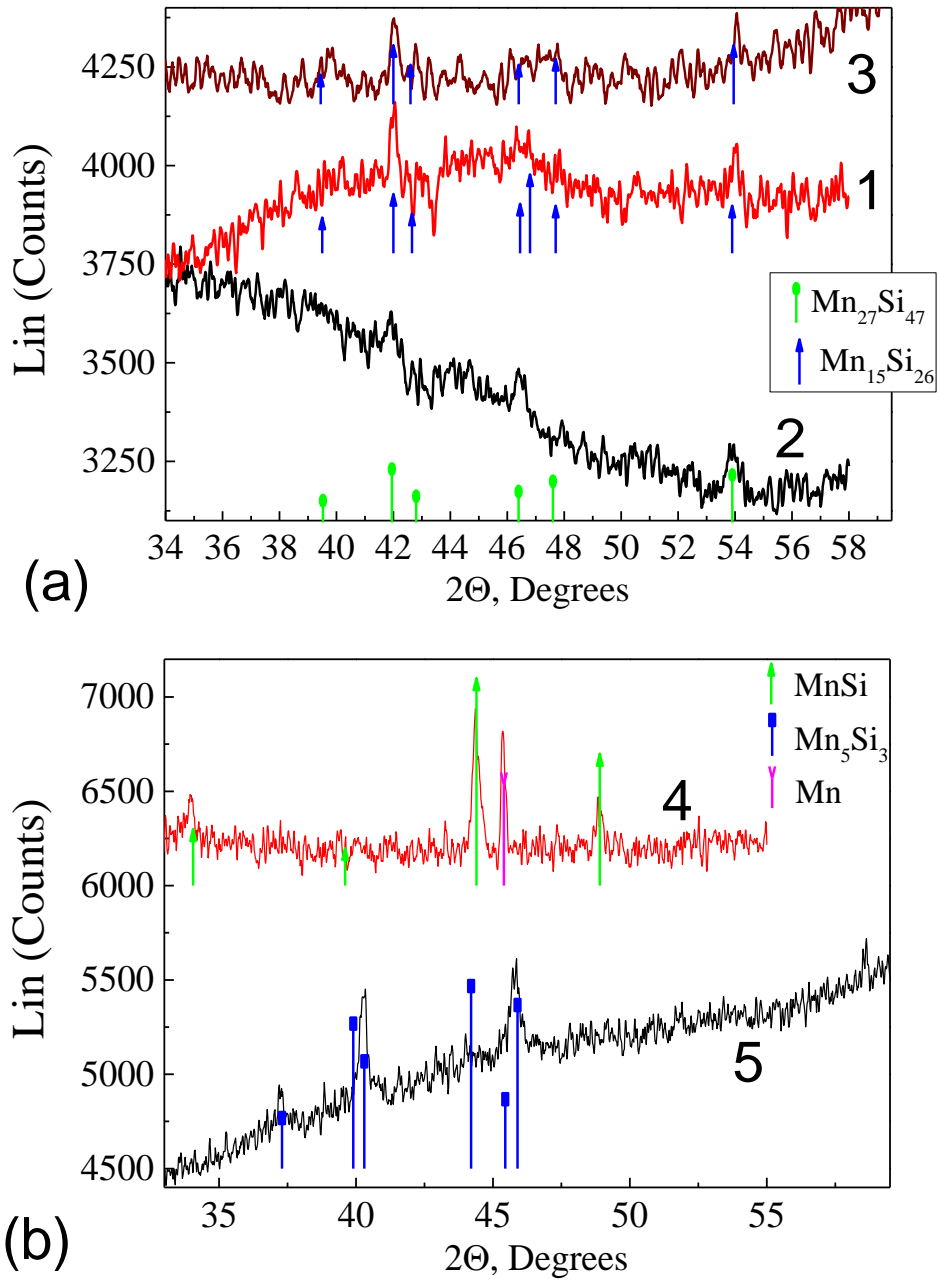

Fig.3 XRD spectra of investigated $Mn_xSi_y$ films. Number of the curve corresponds to number of the sample. Arrows indicate position of diffraction lines for different manganese silicide phases, the latter was derived from databases.

Figure 4 represents the results of XPS chemical analysis: Mn and Si concentration profiles for samples 1 and 4 are depicted. The data for rest of the samples is similar with only difference in the value of concentration. X-axis shows etching time of $Mn_xSi_y$ film, which is proportional to the depth of analysis. The X-axis can be re-calibrated into the distance from surface by dividing etching time by the speed of etching. However, this procedure is unnecessary for the following analysis. Disregarding the data collected on samples surface (which demonstrate the presence of oxides and carbon-based contamination) one can see that films demonstrate uniform depth distribution of elements concentration. The Mn and Si contents were therefore calculated by averaging over 3 etching times. The concentration of Si-bound-to-Mn atoms was obtained in same way and the results are presented in Table 1.

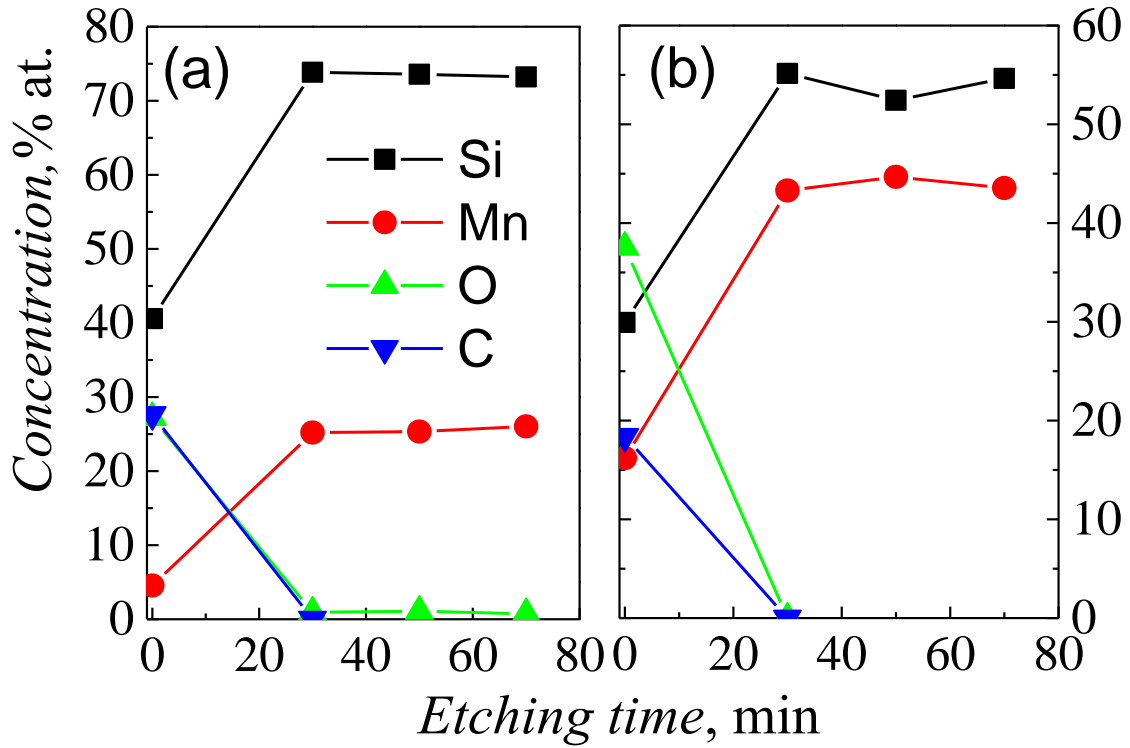

Fig.4 The concentration profiles over the etching time for Si, Mn, O and C measured by XPS technique for sample 1 (a) and sample 4(b)

The chemical analysis can provide some additional information on phase composition of $Mn_xSi_y$ films. Let us introduce the multiplication factor η which is defined as the average number of Si atoms bound to one Mn atom. The value of η can be experimentally estimated from formula

$$\eta = C_{Si-Mn}/C_{Mn}, \quad (1)$$

where $C_{Si-Mn}$ – ratio of Si atoms bound to Mn ($C_{Si-Mn}+C_{Si-Si}=1-C_{Mn}$).

In pure higher manganese silicide films all Si atoms are bound to Mn [41,43]. Considering the composition of $MnSi_{1.75-\delta}$ (where δ≈0-0.02) the multiplication factor in this case would be $\eta=1.75-\delta$. In sample 1, concentration of Si-bound-to-Mn atoms is 2.18 times higher than Mn concentration (η≈2.18). This means that some of Mn atoms are embedded into a different phase with bigger number of Si neighbors. We suppose that at least in low Mn-content samples the possible form of Mn incorporation is Mn doping of a disordered Si phase. This conclusion can be supported by Mn-Si phase diagram, there no stable compound with Mn content below $MnSi_{1.75-\delta}$ is shown [43], and by results obtained within the similar structures [29].

In Mn doped Si phase the exact concentration of Si-bound-to-Mn atoms can vary depending on the exact Mn position: for substitutional Mn η=4 [14], for interstitial Mn η=5-8 [11,14,17], for Mn-Mn complexes η≥3 [11]. The interstitial incorporation of Mn in Si is more favorable as was assumed in [11,14,17]. In view of the possibility of Mn-Mn complex formation we estimate multiplication factor for Si:Mn phase as η=5. In case of sample 1 the multiplication factor is between 1.73 and 5 (η=2.18) and thus one can expect that film includes both higher manganese silicide and Si:Mn phases so the resultant value of η is averaged over these phases.

Let us consider the rest of the samples in such way. In sample 2 the η value is 1.86 which is much closer to that in a higher manganese silicide. Smaller multiplication factor evidences that Mn content in Si:Mn phase is rather low while fraction of higher manganese silicide phase is bigger as compared with sample 1.

In samples with Mn concentration of 33 at. % (sample 3) the multiplication factor is ~1.63, which is below the value suggested for a higher manganese silicide phase. However, the presence of this phase in the film is confirmed by X-Ray diffraction data (Fig.1, Curve 3). The lower value of η as compared with crystalline higher manganese silicide can thus be attributed with the revealed low crystalline quality of the film, which implies the presence of large concentration of defects, such as Mn – rich complexes, or HMS/Si:Mn interfaces. These defects can be responsible for decreasing the number of Si-Mn bonds.

Finally, in samples with high Mn content (4 and 5) all Si atoms are bound to Mn. For this case the analysis of $\eta$ values cannot bring out presence or absence of any of $Mn_5Si_3$, MnSi or Si:Mn phases. For these samples, we can rely only on XRD and relative magnetization data.

## 5. Discussion

Thus four phases were revealed in the investigated films: Si:Mn phase with relatively low Mn content, higher manganese silicide phase, manganese monosilicide and $Mn_5Si_3$ phases. Strong variation of the magnetic properties from sample to sample is believed to be due to the interplay between these different manganese silicide phases. Since higher manganese silicide phase is weakly magnetic (according to all previously obtained data [21,22,34,41,44]), the enhancement of magnetic properties should be attributed with other three phases.

Quite simple reasoning can be performed for the samples with low Mn content. In these samples we can assume the presence of only HMS and Si:Mn phases. First one was revealed by X-Ray diffraction, the second should be present to compensate the lack of Mn. The content of MnSi or $Mn_5Si_3$ should be insignificant since their presence would require more Mn than the presence of HMS and Si:Mn. Then relatively high magnetic signal is attributed with Mn doped silicon phase. In this case magnetization value should depend strongly on Mn concentration in Si:Mn phase.

Room temperature ferromagnetism was earlier demonstrated for Mn doped single-phase silicon films with the lowest Mn content of 0.1 to 0.8 at. %. The samples were obtained by Mn implantation into Si. Hole mediated ferromagnetism mechanism was suggested as in (III.Mn)V diluted magnetic semiconductors [27].

In molecular beam epitaxy (MBE) grown samples with Mn content varying from 0.5 to 1.5 at. % ferromagnetic properties were observed below the room temperature [9]. It was demonstrated that samples were uniformly doped and Mn atoms were in substitutional position of Si lattice. The magnetism was believed to be carrier-mediated, so Curie temperature was increasing with the increase of Mn content.

Further increase of Mn content to 3.6-5 at. % was carried out in [12,16]. It was demonstrated that samples were ferromagnetic with Curie temperatures above 300 K. As in [9] samples were uniformly doped and Curie temperature was monotonically increasing with the increase of Mn content.

Even greater increase of Mn content in Si film as compared to [9,12,16] was performed in [30], there samples were grown by MBE technique. This led to decrease of Curie temperature to the value below 300 K. Structural analysis showed formation of the second phase in Si:Mn films which could probably have led to decrease of Mn concentration in Si:Mn and, as a consequence, to the decrease of Curie temperature. We believe that increase of Mn content performed in [30] triggers higher manganese silicide phase formation, which affects the magnetic properties.

In the investigated samples, Mn content is even higher than that of the films discussed in [30]. From our opinion this leads to further enhancement of higher manganese silicide phase formation whereas both Mn content and fraction of Si:Mn phase decrease. Such reduction is followed by decrease of magnetization since magnetization is believed to be sensitive to Mn content in Si:Mn phase. The reduction of Mn concentration in Si:Mn phase was assumed for sample 2 for which the lowest magnetization value (among ferromagnetic samples) was obtained at room temperature.

In sample 1, on the contrary, both fraction of Si:Mn phase and Mn content in this phase are suggested to be high enough to increase magnetization.

In sample 3 Mn content is about 33,2 % which is close to the higher manganese silicide phase. We believe that in this case most of Mn atoms are located in the HMS phase whereas the fraction of a disordered Si:Mn phase is small. These conditions provide the loss of room temperature magnetism in sample 3.

By analyzing the data on magnetization and phase composition one can make some estimation of Mn concentration in Si:Mn phase. First, we postulate that pure HMS phase is paramagnetic at room temperature and gives no contribution to a saturation magnetization in the range of applied magnetic fields. The magnetization then is attributed with Si:Mn phase. The reported values of magnetization for this phase is varied from 3 to 5 $\mu_b$/Mn [11,20,31,45]. We use value of 4.1 $\mu_b$/Mn since it was experimentally obtained for the films with low Mn content.

The sum of Mn atoms in HMS phase and Si:Mn phase gives the total concentration of Mn:

$$N_1+N_2=C_{Mn} \qquad (2)$$

On the other hand total magnetization values (in Bohr magnetons) is the magnetization of atoms in Si:Mn phase (4.1 times $N_2$). Then Mn content in Si:Mn phase ($N_2$) can be estimated as:

$$N_2=M/N\mu_b \qquad (3)$$

Such analysis gives ~2 % of the entire Mn in Si:Mn phase for sample 1 and ~ 1 % for sample 2. The results are presented in Table 2. We note that such Mn content is sufficient to obtain ferromagnetic properties at room temperature [12,16].

Second, the same values can be derived from the chemical analysis. Again we postulate that the multiplication factor of higher manganese silicide phase is 1.75 whereas in Si:Mn phase $\eta = 5$. In fact, one has to take into account that in disordered HMS film the $\eta$ value can be lower than 1.75 as was revealed for sample 3. The $\eta = 1.75$ would then be the upper limit for the HMS phase. Then the following equation can be written to define the concentration of Si-bound-to-Mn atoms derived from XPS.

$$C_{Si} = 1.75 \times N_1 + 5 \times N_2 \qquad (4)$$

The equation 2 is still valid since it defines the entire Mn content. So far as the values of $C_{Mn}$ and $C_{Si}$ were experimentally found, the solution of equation 2 and 4 allows one to carry out the other evaluation of Mn content in Si:Mn. The results are also given in Table 2.

Table 2. The estimate of Mn atoms concentration in low Mn-content phase of samples 1 and 2

|  | $N_2$, at. % (magnetization) | $N_2$, at. % (chem. analysis) |
|---|---|---|
| Sample 1 | 2.05 | 3.47 |
| Sample 2 | 0.99 | 0.98 |

We note that there is a 1 at. % disagreement between the values obtained from magnetic analysis and chemical analysis of sample 1. However, two estimations performed are in more or less good agreement with each other since there is some degree of freedom in selecting the $\eta$ and average magnetization values for Si:Mn phase which can change the calculated values approximately by half. Thus in samples with Mn content from 24 to 29 at. % magnetism is believed to be driven by the Mn-doped Si phase. Rise of the fraction of this phase and Mn content in it increases the magnetization.

Considering samples 4 and 5 with near 50 at. % Mn concentration chemical analysis cannot be performed, since the multiplication factor is undefined as discussed above. Relative magnetization values for samples 4 and 5 (0.26 and 0.22 $\mu_b$/Mn respectively) are again lower than that in ferromagnetic MnSi [3] and Mn$_5$Si$_3$ [18] phases. This can draw us to conclusion that Mn atoms in MnSi and Mn$_5$Si$_3$ phases are not ordered ferromagnetically and thus the mechanism of the room-temperature ferromagnetism is similar to that of 1-3 samples, i.e. due to the presence of low Mn content magnetic phases in the film. However, no information on the phase composition (besides the dominating phase) can be derived from the experimental results and so magnetization analysis cannot be done.

## 6. Conclusion

In conclusion, we have fabricated and investigated Mn$_x$Si$_y$/i-GaAs(100) thin films with Mn-Si composition varied within the range of 24 to 52 at. %. It was demonstrated that the films revealing room temperature ferromagnetic properties are multiphase systems. One of the phases is a compound with Mn content being higher than the average Mn content of the film (higher manganese silicide, MnSi, Mn$_5$Si$_3$). These phases are believed to be non-ferromagnetic. The other phases are Mn depleted compounds and solid solutions. Room temperature ferromagnetic properties are believed to be due to the low Mn content phases, in particular Mn-doped silicon phase (Si:Mn). Since the formation of these phases is an incidental process, which depends strongly on the Mn/Si flux ratio, the magnetic properties of manganese silicide films are strongly dependent on the technological parameters of the films growth.

**Acknowledgements**

This work was supported by Russian Science Foundation [project no. 17-79-20173].